# UAV-Assisted Weather Radar Calibration: A Theoretical Model for Wind Influence on Metal Sphere Reflectivity

Jiabiao Zhao[1,2], Da Li[1,2], Jiayuan Cui[1,2], Houjun Sun[1,2], Jianjun Ma[1,2]
[1]School of Integrated Circuits and Electronics, Beijing Institute of Technology, Beijing, 100081 China
[2]Beijing Key Laboratory of Millimeter and Terahertz Wave Technology, Beijing, 100081 China

*Abstract*- **The calibration of weather radar for detecting meteorological phenomena has advanced rapidly, aiming to enhance accuracy. Utilizing an unmanned aerial vehicle (UAV) equipped with a suspended metal sphere introduces an efficient calibration method by allowing dynamic adjustment of the UAV's position, effectively acting as a mobile calibration platform. However, external factors such as wind can introduce bias in reflectivity measurements by causing the sphere to deviate from its intended position. This study develops a theoretical model to assess the impact of the metal sphere's one-dimensional oscillation on reflectivity. The findings offer valuable insights for UAV-based radar calibration efforts.**

## I. Introduction

Weather radar plays a pivotal role in atmospheric science by providing essential data on cloud dynamics, hydrometeor distribution, and precipitation severity [1]. Consequently, it is indispensable for studying weather and climate dynamics, evaluating cloud system models, and forecasting hazardous precipitation events [2]. The accuracy of radar-based weather detection critically depends on proper calibration, which ensures the reliability of radar data for scientific and operational purposes. Traditional calibration methods include the drizzle [3], sun [4], and test signal techniques [5], each aiming to minimize reflectivity errors across various radar types by correcting system imperfections. Employing a standard metal sphere as a calibration target is a well-established method [6]. Given the advancements in unmanned aerial vehicle (UAV) technology, using UAVs to deploy metal spheres for calibration offers significant benefits in terms of positioning accuracy, flexibility, and operational speed. Nevertheless, calibration accuracy can be compromised by external disturbances like wind, which can displace the metal sphere from its target area. It should be mentioned that the calibration mentioned earlier only focus on the methods, lacking the research on the effects of external environment on calibration.

In this work, we present a theoretical model to simulate the effect of wind on the reflectivity measurements obtained during UAV-assisted calibration processes.

## II. Theoretical Modeling

### A. Calibration theory of metal sphere

Figure. 1 shows the schematic of calibration for UAV with the suspended metal sphere. For a detection target with a known radar cross section (RCS) $\sigma$, according to the radar equation[7]:

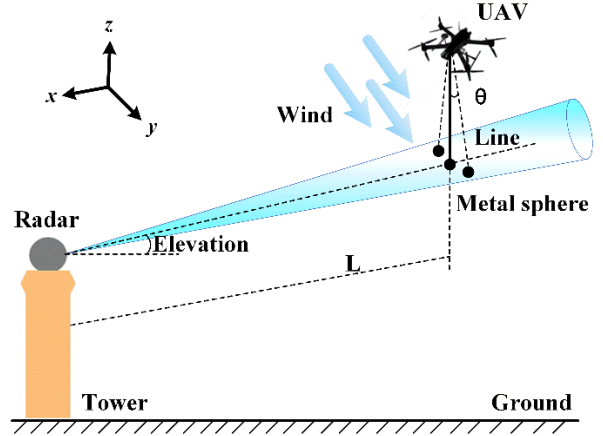

Figure 1. Schematic of calibration for UAV with suspended metal sphere.

$$P_r = \frac{P_t G^2 \lambda^2 \sigma}{(4\pi)^3 R^4} \quad (1)$$

where $P_r$ is the echo power of the target received by the radar, $P_t$ is the peak power of the radar transmitting pulse, $G$ is the antenna gain, $\lambda$ is the radar wavelength, $R$ is the distance from the radar to the target. When the backscattering from the metal sphere lies in the region of geometrical optics and its radius $r$ satisfies the condition:

$$\frac{2\pi r}{\lambda} \gg 1 \quad (2)$$

Then, the RCS of metal sphere depends on the $r$, i.e., $\sigma \sim \pi r^2$. Combined with radar system parameters, the reflectivity of metal sphere can be written as

$$Z = C + P_r + 20 \lg R + RA_t \quad (3)$$

where C is radar constant, $A_t$ is atmospheric loss. During the calibration process, ensure that the $R$ meets the far-field condition. When the radar detects the metal sphere, it must continuously adjust the metal sphere position to optimize the reflected signal to be in the direction of the main lobe of the antenna. The RCS of metal sphere is known, and the ideal reflectivity of metal sphere could be calculated. Finally, the actual reflectivity is compared with the ideal reflectivity.

## B. Analysis of calibration errors

The UAV with the suspended metal sphere can be regarded as a mobile calibration tower, which increases the calibration efficiency. Calibration is inevitably affected by the external winds. For simplicity but for reference, the radar beam is perpendicular to the $yoz$ plane and the movement of the metal sphere is one-dimensional oscillation around the UAV, represented in Fig. (1) as perpendicular to the $xoz$ plane. The angle of oscillation around the UAV is $\theta$, which ranges from -10° to 10°. The length of line is 60 m and the elevation of radar is 2°. The angle of the metal sphere deviates from the main lobe of the antenna (the direction of the maximum power) is $\beta$. As shown in Fig. 2(a), when the straight-line distance $L$ ranges from 2 km to 5 km, the variation of $\beta$ does not exceed 0.3°. While the detected target deviates from the direction of the normal beam of antenna, the equivalent RCS can be approximately expressed as

$$\sigma' = \sigma \left| \sin c \left( \frac{\beta}{\beta_{ant.}/2} \right) \right| \qquad (4)$$

where $\beta_{ant.}$ is the beamwidth of antenna. The variation of $\sigma'$ is shown in Fig. 2(b), overall, the value of $\sigma'$ is not more than 0.1 m². Since longer L means wider 3dB beam-range, as a result, a lower gain loss for the radar receiving antennas. Fig. 2(c) and Fig. 2(d) report the reflectivity $Z$ and $\Delta Z$ of metal sphere with different $\theta$. Since the reflectivity is usually expressed in units of $mm^6 m^{-3}$ and the interval of variation in Z-values can vary over many orders of magnitude, the logarithmic transformation $10 \times log10(Z)$ is utilized, with its units being in decibels of $Z$ [8]. From Fig. 2(c) and Fig. 2(d), it can be concluded that longer $L$ leads to smaller deviation angle with respect to the antenna normal beam. When $L$ varies from 2 km to 5 km, the $\Delta Z$ is reduced from 5.9 dB to 0.82 dB, Therefore, the calibration error can be easily calculated when the straight-line distance is relatively far from the radar. It should be pointed out that the calibration of reflectivity is rarely affected (0.01dB) by the metal sphere of different materials (i.e., tinfoil and metal) [7]. In addition, it demonstrates that radar constants are consistent when the calibration experiments use spheres of different sizes, all of which are related to the Mie scattering region [9].

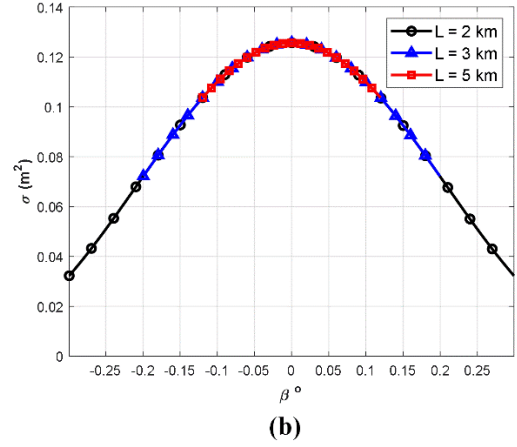

(b)

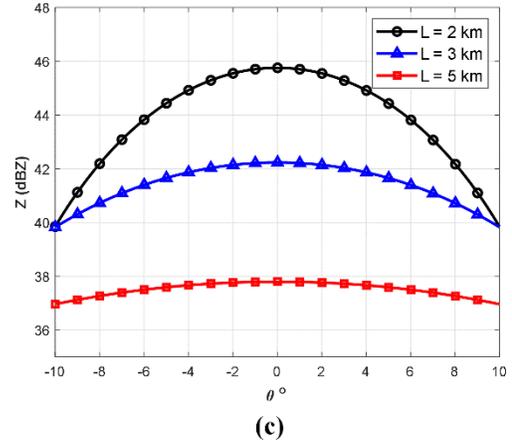

(c)

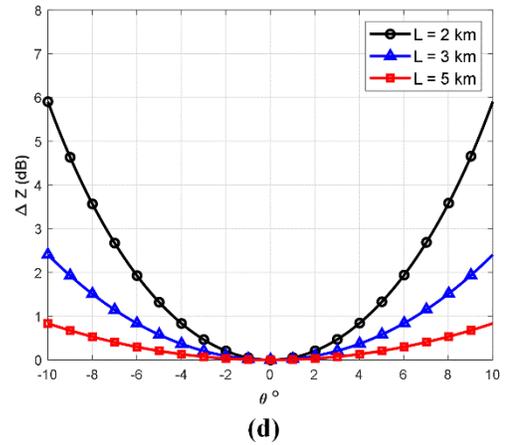

(d)

Figure 2. Schematic of variation of the straight-line distance $L$. (a) Variation of $\beta$ with respect to the $L$. (b) Variation of $\sigma$ with respect to the $\beta$. (c) Variation of $Z$ with respect to the $\theta$. (d) Variation of $\Delta Z$ with respect to the $\theta$.

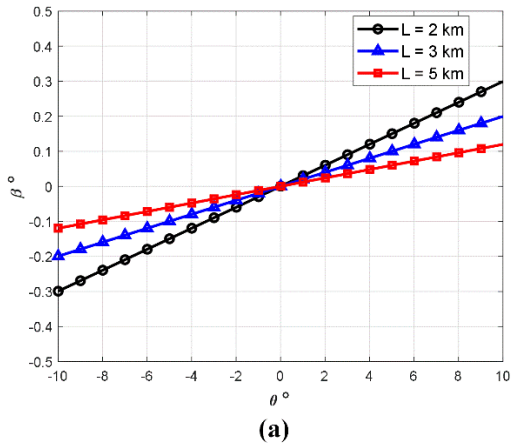

(a)

## III. CONCLUSION

In this work, we introduce a theoretical model to evaluate the impact of wind on the reflectivity of a UAV-mounted metal sphere used in radar calibration. By extending the operational range, the adverse effects of oscillation on reflectivity measurement accuracy can be minimized, facilitating more precise calibration. This model serves as a guide for optimizing UAV-based radar calibration processes. In the near future, we propose to take the measurements and modeling by including the trajectories of UAV.